%% file: main.tex
\documentclass[sigconf]{acmart}

\usepackage{soul, color}
\usepackage{mdframed}
\usepackage{tikz}
\usepackage{subcaption}
\usetikzlibrary{calc,positioning,shapes.geometric,shapes.symbols,shapes.misc}
\usepackage{bm}

\AtBeginDocument{%
  \providecommand\BibTeX{{%
    \normalfont B\kern-0.5em{\scshape i\kern-0.25em b}\kern-0.8em\TeX}}}

\setcopyright{acmcopyright}
\copyrightyear{2023}
\acmYear{2023}
\acmDOI{XXXXXXX.XXXXXXX}

\acmConference[CONSEQUENCES '24]{Causality, Counterfactuals \& Sequential Decision-Making}{Oct. 14th-18th 2024}{Bari, Italy}

\acmBooktitle{CONSEQUENCES '24: Causality, Counterfactuals \& Sequential Decision-Making, Oct. 14th-18th 2024, Bari, Italy} 

\begin{document}

\title[Causal Discovery in Recommender Systems]{Causal Discovery in Recommender Systems: Example and Discussion}

\author{Emanuele Cavenaghi}
\email{ecavenaghi@unibz.it}
\orcid{0000-0002-0235-0421}
\affiliation{%
    \institution{Free University of Bozen-Bolzano}
    \streetaddress{Piazza Università, 1}
    \city{Bolzano}
    \country{Italy}
    \postcode{39100}
}
\author{Fabio Stella}
\email{fabio.stella@unimib.it}
\orcid{0000-0002-1394-0507}
\affiliation{%
    \institution{University of Milano-Bicocca}
    \streetaddress{Piazza dell'Ateneo Nuovo, 1}
    \city{Milano}
    \country{Italy}
    \postcode{20126}
}
\author{Markus Zanker}
\email{mzanker@unibz.it}
\orcid{0000-0002-4805-5516}
\affiliation{%
    \institution{Free University of Bozen-Bolzano}
    \streetaddress{Piazza Università, 1}
    \city{Bolzano}
    \country{Italy}
    \postcode{39100}
}
\affiliation{%
    \institution{University of Klagenfurt}
    \streetaddress{Universitätsstraße 65-67}
    \city{Klagenfurt}
    \country{Austria}
    \postcode{9020}
}

\begin{abstract}
    Causality is receiving increasing attention by the artificial intelligence and machine learning communities. This paper gives an example of modelling a recommender system problem using causal graphs. Specifically, we approached the causal discovery task to learn a causal graph by combining observational data from an open-source dataset with prior knowledge. The resulting causal graph shows that only a few variables effectively influence the analysed feedback signals. This contrasts with the recent trend in the machine learning community to include more and more variables in massive models, such as neural networks.     
\end{abstract}


\maketitle

\input{sections/01_introduction}
\input{sections/02_causal_discovery_process}
\input{sections/03_learned_causal_graph}

\input{sections/04_conclusions}

\bibliographystyle{ACM-Reference-Format}
\bibliography{references}

\newpage
\appendix

\section{Full Learned Causal Graph} \label{sec:appendix_full_cg}

\begin{figure*}
    \centering
    \includegraphics[width=\textwidth]{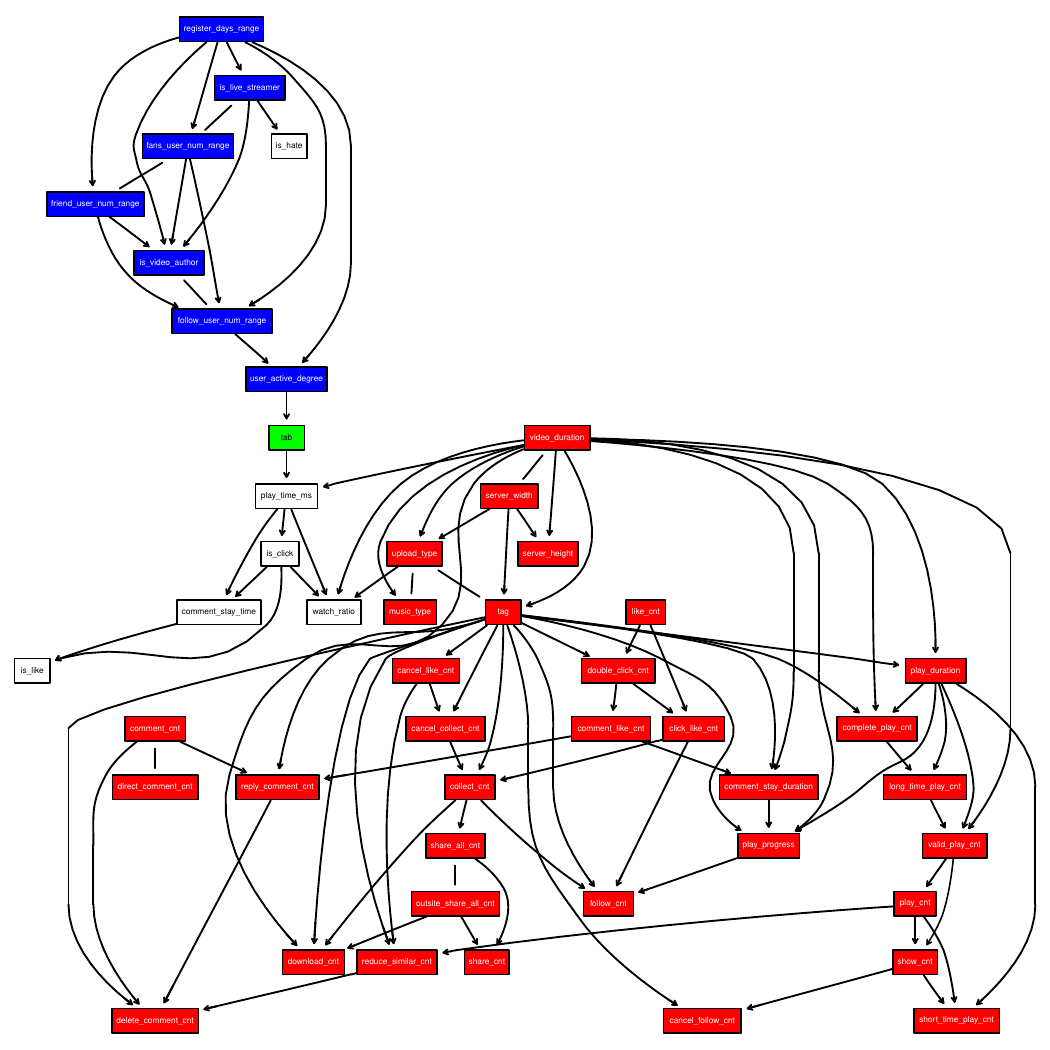}
    \caption{Full SCM learned from data and expert knowledge.}
    \label{fig:full_cg}
\end{figure*}

\end{document}

%% file: sections/01_introduction.tex
\section{Introduction}

In recent years, several approaches from causality have been applied in \textit{Recommender Systems} (RS) \cite{chen2023bias, gao2022causal, cavenaghi2023towards}, especially using Causal Graphs (CGs) \cite{pearl2009causality} which allow us to model, in a graphical and human-readable format, cause-and-effect relations among factors. Furthermore, to exploit the strength of causality, we should define the quantity that we aim to estimate as \textit{causal estimands}\footnote{A causal estimand is a quantity that encodes the notion of the causal effect of a variable (the cause) on another (the effect).} that can only be estimated in controlled experiments. However, through CGs, we can identify an equivalent \textit{statistical estimand} that can be estimated using observational data, i.e., data collected outside the scope of controlled experiments. To this end, a common approach is to use the so-called \textit{adjustment formula} estimator \cite{glymour2016causal} on an \textit{adjustment set} identified on the CG. Moreover, even the commonly used causal approaches, such as \textit{Inverse Probability Weighting} \cite{schnabel2016recommendations, wang2016learning, joachims2017unbiased}, doubly robust estimators \cite{saito2020doubly} and counterfactuals \cite{wang2021clicks, wei2021model, yang2021top}, can benefit from the use of CGs to identify the relevant variables that should be used and thus reduce the dimension of the problem. 

However, very few works tackle the problem of learning a CG in an RS domain \cite{xu2022causal, cavenaghi2023causal} from data. Nonetheless, a manually designed CG only encodes our knowledge of the problem without any information gathered from the data. Therefore, we tackled the problem of learning a CG by combining prior knowledge and observational data from the open-source dataset \textit{KuaiRand} \cite{gao2022kuairand}. As a result, we report all the steps followed to learn a CG, from data and prior knowledge to the final result, to inspire other researchers who want to use causal models to tackle RS problems.

%% file: sections/02_causal_discovery_process.tex
\section{Causal Discovery Process}

In this section, we report the process followed to learn the CG reported below starting from the open-source dataset \textit{KuaiRand} \cite{gao2022kuairand}. In particular, we used the \textit{KuaiRand-Pure} dataset collected through randomly recommended videos in the user interaction sequence. This version of the dataset consists of $1\,186\,059$ interactions between the system and the $27\,285$ users who were recommended from a pool of $7\,583$ items. Moreover, the dataset includes $30$ users' features, $62$ items' features and $12$ feedback signals. Below, we report the five macro-steps followed in the causal discovery process:

\paragraph{Remove features}
In the first step, we removed the interaction features irrelevant to our goal, such as the interaction date and time, and the feedback signals with too few positive observations. Furthermore, we excluded all the encrypted users' features since we can not use prior knowledge on them as they lack any semantic information. Finally, we removed the items' music ID since almost all the videos are related to a different ID.

\paragraph{Discretize features}
In the second step, we discretized all the features since the structure learning algorithm used to learn the CG can only handle discrete data. To this end, several features are available both with the original values and with a discretized version; thus, for these features, we decided to keep the discretization suggested by the authors of the dataset. For the remaining features, we chose the categories using the feature semantics to lose as little information as possible and to ensure that each category was observed a reasonable number of times. 

\paragraph{Build Prior-Knowledge}
In the third step, we defined how to include prior knowledge in the structure learning phase. To this extent, domain experts can list specific edges that the structure learning algorithm must exclude or include during the graph recovering procedure, i.e., forbidden and required edges. Alternatively, it is also possible to define ordered edge sets, or tiers, that induce a partial order among the observed variables where nodes in lower tiers can not be the cause of nodes in higher tiers. In this work, we used the tiers approach and defined the following tiers: (i) user features, (ii) context feature ``\textit{tab}''\footnote{As named in the dataset and described at https://kuairand.com/}, (iii) item features, (iv) item statistics and (v) feedback signals.

\paragraph{Structure Learning Phase}
In this work, we relied on the \emph{Hill-Climbing} (HC) \cite{scutari2019learning} algorithm implemented in \textit{bnlearn}\footnote{https://www.bnlearn.com/}, which traverses the space of the possible CGs selecting the optimal graph $\mathcal{G}^*$ w.r.t. a goodness-of-fit function $\mathcal{S}$, known as the \textit{scoring criterion}. At its core, HC iteratively modifies the current recovered graph to maximize $\mathcal{S}$ by adding, deleting or reversing individual edges. When no modification improves the score, the procedure halts and returns the current solution. HC is guaranteed to include edges coherent with the underlying independence statements, provided that $\mathcal{S}$ is a \emph{consistent scoring criterion}. In this work, we used the \textit{Bayesian Information Criterion} (BIC) score \cite{chickering1995transformational}.

\paragraph{Average Causal Graph}
In total, we learnt $100$ CGs using the HC algorithm with the prior knowledge and the discretized dataset described above. Then, to obtain a single CG containing only significant edges, we averaged the learnt CGs by selecting only the edges present in at least $90\%$ of the CGs. The learned CG is reported in Appendix \ref{sec:appendix_full_cg} Figure \ref{fig:full_cg}.

%% file: sections/03_learned_causal_graph.tex
\section{Learned Causal Graph}
\begin{figure}[ht]
    \centering
    \includegraphics[width=0.4\textwidth]{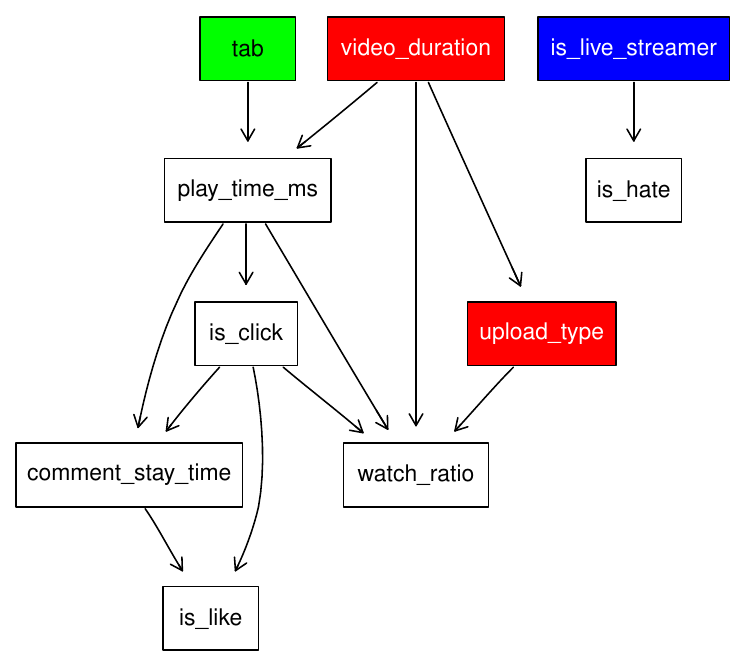}
    \caption{Markov blanket of the feedback signals in the CG. Nodes are coloured by their semantics: users, items and context features are in blue, red and green, respectively.}
    \label{fig:markov_blanket_cg}
\end{figure}

In this section, we report a part of the entire CG leveraging the independence statements to restrict the CG to a proper sub-graph by discarding irrelevant factors that are not in the \textit{Markov Blanket} (MB) \cite{pearl1988probabilistic} of the feedback signals nodes. An MB is defined as a subset of variables that contains all the useful information to infer the value of a random variable. Therefore, since the most important factors are the feedback signals, we restrict our focus on their MB, which is reported in Figure \ref{fig:markov_blanket_cg}.

The first natural questions that arise are: Is the learned CG correct? Why is it different from other models? The answer is closely linked to the primary purpose of causal models. Citing the words of \textit{Judea Pearl} \cite{pearl2014interpretation}: ``The purpose of the diagram is to provide an unambiguous description of the scientific context of a given application''. This \textit{unambiguous description} must not be taken as a dogma but should be used to build a ``hypothetical consensus on what is plausible and important versus that which is deemed negligible or implausible'' \cite{pearl2014interpretation}. This consensus, in the form of possible associations among measured and unmeasured variables, must be built through ``[...] scientific considerations and debated by experts in the field'' \cite{pearl2014interpretation}. Indeed, the correctness of a CG can not be guaranteed by automatic tests/procedures but must be verified through the mix of domain knowledge and observations/experiments. However, this limitation should not be confused with a weakness w.r.t. other models, as no existing model can be guaranteed to be correct. 

Nevertheless, unlike other models, experts can debate the cause-and-effect relations and dependencies/independencies embedded in a CG to build the \textit{hypothetical consensus} mentioned above. For example, according to the sub-CG in Figure \ref{fig:markov_blanket_cg}, the only relevant items' features that influence the feedback the users give are the video duration and the upload type, while all the others are irrelevant. Therefore, following this model, our recommendations should consider only these two variables, disregarding all the others to avoid introducing noise in the decision and drastically reducing the amount of data required. Similarly, the analysed users' features seem relevant only for the \textit{is\_hate} feedback. However, in this case, we should note that the available features only describe the user on the platform, while the ones that identify a user are encrypted.

%% file: sections/04_conclusions.tex
\section{Conclusions}

This work focused on the causal discovery task of learning a CG by combining observational data from the KuaiRand open-source dataset with prior knowledge. The resulting CG contrasts the recent trend in Machine Learning and RSs to include more and more variables in larger and larger models. Indeed, the learned CG suggests that only a few variables effectively influence the analysed feedback signals. Therefore, all the others are irrelevant to our decisions and only contribute to noise. Moreover, more data is required as the number of variables (i.e., dimensions) increases. This finding can be supported by the fact that users (i.e., people) can not consider many factors when deciding, and their decisions are usually based on a few essential factors and subconscious emotional feelings \cite{kahneman2021noise}. On the other hand, part of the effect of the recommended items is (usually) not captured by the available factors. Therefore, more emphasis should be placed on modelling the problem and collecting the necessary information to make informed decisions.

%% file: main.bbl

\begin{thebibliography}{20}


\ifx \showCODEN    \undefined \def \showCODEN     #1{\unskip}     \fi
\ifx \showDOI      \undefined \def \showDOI       #1{#1}\fi
\ifx \showISBNx    \undefined \def \showISBNx     #1{\unskip}     \fi
\ifx \showISBNxiii \undefined \def \showISBNxiii  #1{\unskip}     \fi
\ifx \showISSN     \undefined \def \showISSN      #1{\unskip}     \fi
\ifx \showLCCN     \undefined \def \showLCCN      #1{\unskip}     \fi
\ifx \shownote     \undefined \def \shownote      #1{#1}          \fi
\ifx \showarticletitle \undefined \def \showarticletitle #1{#1}   \fi
\ifx \showURL      \undefined \def \showURL       {\relax}        \fi
\providecommand\bibfield[2]{#2}
\providecommand\bibinfo[2]{#2}
\providecommand\natexlab[1]{#1}
\providecommand\showeprint[2][]{arXiv:#2}

\bibitem[Cavenaghi et~al\mbox{.}(2023)]%
        {cavenaghi2023causal}
\bibfield{author}{\bibinfo{person}{Emanuele Cavenaghi}, \bibinfo{person}{Alessio Zanga}, \bibinfo{person}{A Rimoldi}, \bibinfo{person}{P Minasi}, \bibinfo{person}{F Stella}, \bibinfo{person}{M Zanker}, {et~al\mbox{.}}} \bibinfo{year}{2023}\natexlab{}.
\newblock \showarticletitle{Causal Discovery in Recommender Systems: a Case Study in Online Hotel Search}.
\newblock  (\bibinfo{year}{2023}).
\newblock


\bibitem[Cavenaghi et~al\mbox{.}(2024)]%
        {cavenaghi2023towards}
\bibfield{author}{\bibinfo{person}{Emanuele Cavenaghi}, \bibinfo{person}{Alessio Zanga}, \bibinfo{person}{Fabio Stella}, {and} \bibinfo{person}{Markus Zanker}.} \bibinfo{year}{2024}\natexlab{}.
\newblock \showarticletitle{Towards a Causal Decision-Making Framework for Recommender Systems}.
\newblock \bibinfo{journal}{\emph{ACM Trans. Recomm. Syst.}} \bibinfo{volume}{2}, \bibinfo{number}{2}, Article \bibinfo{articleno}{17} (\bibinfo{year}{2024}), \bibinfo{numpages}{34}~pages.
\newblock


\bibitem[Chen et~al\mbox{.}(2023)]%
        {chen2023bias}
\bibfield{author}{\bibinfo{person}{Jiawei Chen}, \bibinfo{person}{Hande Dong}, \bibinfo{person}{Xiang Wang}, \bibinfo{person}{Fuli Feng}, \bibinfo{person}{Meng Wang}, {and} \bibinfo{person}{Xiangnan He}.} \bibinfo{year}{2023}\natexlab{}.
\newblock \showarticletitle{Bias and Debias in Recommender System: A Survey and Future Directions}.
\newblock \bibinfo{journal}{\emph{ACM Trans. Inf. Syst.}} \bibinfo{volume}{41}, \bibinfo{number}{3}, Article \bibinfo{articleno}{67} (\bibinfo{year}{2023}), \bibinfo{numpages}{39}~pages.
\newblock


\bibitem[Chickering(1995)]%
        {chickering1995transformational}
\bibfield{author}{\bibinfo{person}{David~Maxwell Chickering}.} \bibinfo{year}{1995}\natexlab{}.
\newblock \showarticletitle{A transformational characterization of equivalent Bayesian network structures}. In \bibinfo{booktitle}{\emph{Proceedings of the Eleventh Conference on Uncertainty in Artificial Intelligence}} (Montr\'{e}al, Qu\'{e}, Canada) \emph{(\bibinfo{series}{UAI'95})}. \bibinfo{publisher}{Morgan Kaufmann Publishers Inc.}, \bibinfo{address}{San Francisco, CA, USA}, \bibinfo{pages}{87–98}.
\newblock


\bibitem[Gao et~al\mbox{.}(2022a)]%
        {gao2022kuairand}
\bibfield{author}{\bibinfo{person}{Chongming Gao}, \bibinfo{person}{Shijun Li}, \bibinfo{person}{Yuan Zhang}, \bibinfo{person}{Jiawei Chen}, \bibinfo{person}{Biao Li}, \bibinfo{person}{Wenqiang Lei}, \bibinfo{person}{Peng Jiang}, {and} \bibinfo{person}{Xiangnan He}.} \bibinfo{year}{2022}\natexlab{a}.
\newblock \showarticletitle{KuaiRand: An Unbiased Sequential Recommendation Dataset with Randomly Exposed Videos}. In \bibinfo{booktitle}{\emph{Proceedings of the 31st ACM International Conference on Information and Knowledge Management}} (Atlanta, GA, USA) \emph{(\bibinfo{series}{CIKM '22})}. \bibinfo{pages}{3953–3957}.
\newblock


\bibitem[Gao et~al\mbox{.}(2022b)]%
        {gao2022causal}
\bibfield{author}{\bibinfo{person}{Chen Gao}, \bibinfo{person}{Yu Zheng}, \bibinfo{person}{Wenjie Wang}, \bibinfo{person}{Fuli Feng}, \bibinfo{person}{Xiangnan He}, {and} \bibinfo{person}{Yong Li}.} \bibinfo{year}{2022}\natexlab{b}.
\newblock \showarticletitle{Causal Inference in Recommender Systems: A Survey and Future Directions}.
\newblock \bibinfo{journal}{\emph{arXiv preprint arXiv:2208.12397}}  \bibinfo{volume}{1} (\bibinfo{year}{2022}), \bibinfo{numpages}{29}~pages.
\newblock


\bibitem[Glymour et~al\mbox{.}(2016)]%
        {glymour2016causal}
\bibfield{author}{\bibinfo{person}{Madelyn Glymour}, \bibinfo{person}{Judea Pearl}, {and} \bibinfo{person}{Nicholas~P Jewell}.} \bibinfo{year}{2016}\natexlab{}.
\newblock \bibinfo{booktitle}{\emph{Causal inference in statistics: A primer}}.
\newblock \bibinfo{publisher}{John Wiley \& Sons}, \bibinfo{address}{Hoboken, United States}.
\newblock


\bibitem[Joachims et~al\mbox{.}(2017)]%
        {joachims2017unbiased}
\bibfield{author}{\bibinfo{person}{Thorsten Joachims}, \bibinfo{person}{Adith Swaminathan}, {and} \bibinfo{person}{Tobias Schnabel}.} \bibinfo{year}{2017}\natexlab{}.
\newblock \showarticletitle{Unbiased Learning-to-Rank with Biased Feedback}. In \bibinfo{booktitle}{\emph{Proceedings of the Tenth ACM International Conference on Web Search and Data Mining}} (Cambridge, United Kingdom) \emph{(\bibinfo{series}{WSDM '17})}. \bibinfo{publisher}{Association for Computing Machinery}, \bibinfo{address}{New York, NY, USA}, \bibinfo{pages}{781–789}.
\newblock


\bibitem[Kahneman et~al\mbox{.}(2021)]%
        {kahneman2021noise}
\bibfield{author}{\bibinfo{person}{Daniel Kahneman}, \bibinfo{person}{Olivier Sibony}, {and} \bibinfo{person}{Cass~R Sunstein}.} \bibinfo{year}{2021}\natexlab{}.
\newblock \bibinfo{booktitle}{\emph{Noise: A flaw in human judgment}}.
\newblock \bibinfo{publisher}{Hachette UK}.
\newblock


\bibitem[Pearl(1988)]%
        {pearl1988probabilistic}
\bibfield{author}{\bibinfo{person}{Judea Pearl}.} \bibinfo{year}{1988}\natexlab{}.
\newblock \bibinfo{booktitle}{\emph{Probabilistic reasoning in intelligent systems: networks of plausible inference}}.
\newblock \bibinfo{publisher}{Morgan kaufmann}, \bibinfo{address}{Cambridge, Massachusetts, United States}.
\newblock


\bibitem[Pearl(2009)]%
        {pearl2009causality}
\bibfield{author}{\bibinfo{person}{Judea Pearl}.} \bibinfo{year}{2009}\natexlab{}.
\newblock \bibinfo{booktitle}{\emph{Causality}}.
\newblock \bibinfo{publisher}{Cambridge university press}, \bibinfo{address}{Cambridge, United Kingdom}.
\newblock


\bibitem[Pearl(2014)]%
        {pearl2014interpretation}
\bibfield{author}{\bibinfo{person}{Judea Pearl}.} \bibinfo{year}{2014}\natexlab{}.
\newblock \showarticletitle{Interpretation and identification of causal mediation.}
\newblock \bibinfo{journal}{\emph{Psychological methods}} \bibinfo{volume}{19}, \bibinfo{number}{4} (\bibinfo{year}{2014}), \bibinfo{pages}{459}.
\newblock


\bibitem[Saito(2020)]%
        {saito2020doubly}
\bibfield{author}{\bibinfo{person}{Yuta Saito}.} \bibinfo{year}{2020}\natexlab{}.
\newblock \showarticletitle{Doubly Robust Estimator for Ranking Metrics with Post-Click Conversions}. In \bibinfo{booktitle}{\emph{Proceedings of the 14th ACM Conference on Recommender Systems}} (Virtual Event, Brazil) \emph{(\bibinfo{series}{RecSys '20})}. \bibinfo{publisher}{Association for Computing Machinery}, \bibinfo{address}{New York, NY, USA}, \bibinfo{pages}{92–100}.
\newblock


\bibitem[Schnabel et~al\mbox{.}(2016)]%
        {schnabel2016recommendations}
\bibfield{author}{\bibinfo{person}{Tobias Schnabel}, \bibinfo{person}{Adith Swaminathan}, \bibinfo{person}{Ashudeep Singh}, \bibinfo{person}{Navin Chandak}, {and} \bibinfo{person}{Thorsten Joachims}.} \bibinfo{year}{2016}\natexlab{}.
\newblock \showarticletitle{Recommendations as Treatments: Debiasing Learning and Evaluation}. In \bibinfo{booktitle}{\emph{Proceedings of The 33rd International Conference on Machine Learning}} \emph{(\bibinfo{series}{Proceedings of Machine Learning Research}, Vol.~\bibinfo{volume}{48})}, \bibfield{editor}{\bibinfo{person}{Maria~Florina Balcan} {and} \bibinfo{person}{Kilian~Q. Weinberger}} (Eds.). \bibinfo{publisher}{PMLR}, \bibinfo{address}{New York, New York, USA}, \bibinfo{pages}{1670--1679}.
\newblock


\bibitem[Scutari et~al\mbox{.}(2019)]%
        {scutari2019learning}
\bibfield{author}{\bibinfo{person}{Marco Scutari}, \bibinfo{person}{Claudia Vitolo}, {and} \bibinfo{person}{Allan Tucker}.} \bibinfo{year}{2019}\natexlab{}.
\newblock \showarticletitle{Learning Bayesian networks from big data with greedy search: computational complexity and efficient implementation}.
\newblock \bibinfo{journal}{\emph{Statistics and Computing}}  \bibinfo{volume}{29} (\bibinfo{year}{2019}), \bibinfo{pages}{1095--1108}.
\newblock


\bibitem[Wang et~al\mbox{.}(2021)]%
        {wang2021clicks}
\bibfield{author}{\bibinfo{person}{Wenjie Wang}, \bibinfo{person}{Fuli Feng}, \bibinfo{person}{Xiangnan He}, \bibinfo{person}{Hanwang Zhang}, {and} \bibinfo{person}{Tat-Seng Chua}.} \bibinfo{year}{2021}\natexlab{}.
\newblock \showarticletitle{Clicks Can Be Cheating: Counterfactual Recommendation for Mitigating Clickbait Issue}. In \bibinfo{booktitle}{\emph{Proceedings of the 44th International ACM SIGIR Conference on Research and Development in Information Retrieval}} (Virtual Event, Canada) \emph{(\bibinfo{series}{SIGIR '21})}. \bibinfo{publisher}{Association for Computing Machinery}, \bibinfo{address}{New York, NY, USA}, \bibinfo{pages}{1288–1297}.
\newblock


\bibitem[Wang et~al\mbox{.}(2016)]%
        {wang2016learning}
\bibfield{author}{\bibinfo{person}{Xuanhui Wang}, \bibinfo{person}{Michael Bendersky}, \bibinfo{person}{Donald Metzler}, {and} \bibinfo{person}{Marc Najork}.} \bibinfo{year}{2016}\natexlab{}.
\newblock \showarticletitle{Learning to Rank with Selection Bias in Personal Search}. In \bibinfo{booktitle}{\emph{Proceedings of the 39th International ACM SIGIR Conference on Research and Development in Information Retrieval}} (Pisa, Italy) \emph{(\bibinfo{series}{SIGIR '16})}. \bibinfo{publisher}{Association for Computing Machinery}, \bibinfo{address}{New York, NY, USA}, \bibinfo{pages}{115–124}.
\newblock


\bibitem[Wei et~al\mbox{.}(2021)]%
        {wei2021model}
\bibfield{author}{\bibinfo{person}{Tianxin Wei}, \bibinfo{person}{Fuli Feng}, \bibinfo{person}{Jiawei Chen}, \bibinfo{person}{Ziwei Wu}, \bibinfo{person}{Jinfeng Yi}, {and} \bibinfo{person}{Xiangnan He}.} \bibinfo{year}{2021}\natexlab{}.
\newblock \showarticletitle{Model-Agnostic Counterfactual Reasoning for Eliminating Popularity Bias in Recommender System}. In \bibinfo{booktitle}{\emph{Proceedings of the 27th ACM SIGKDD Conference on Knowledge Discovery and; Data Mining}} (Virtual Event, Singapore) \emph{(\bibinfo{series}{KDD '21})}. \bibinfo{publisher}{Association for Computing Machinery}, \bibinfo{address}{New York, NY, USA}, \bibinfo{pages}{1791–1800}.
\newblock


\bibitem[Xu et~al\mbox{.}(2022)]%
        {xu2022causal}
\bibfield{author}{\bibinfo{person}{Shuyuan Xu}, \bibinfo{person}{Da Xu}, \bibinfo{person}{Evren Korpeoglu}, \bibinfo{person}{Sushant Kumar}, \bibinfo{person}{Stephen Guo}, \bibinfo{person}{Kannan Achan}, {and} \bibinfo{person}{Yongfeng Zhang}.} \bibinfo{year}{2022}\natexlab{}.
\newblock \showarticletitle{Causal Structure Learning with Recommendation System}.
\newblock \bibinfo{journal}{\emph{arXiv preprint arXiv:2210.10256}}  \bibinfo{volume}{1} (\bibinfo{year}{2022}), \bibinfo{numpages}{10}~pages.
\newblock


\bibitem[Yang et~al\mbox{.}(2021)]%
        {yang2021top}
\bibfield{author}{\bibinfo{person}{Mengyue Yang}, \bibinfo{person}{Quanyu Dai}, \bibinfo{person}{Zhenhua Dong}, \bibinfo{person}{Xu Chen}, \bibinfo{person}{Xiuqiang He}, {and} \bibinfo{person}{Jun Wang}.} \bibinfo{year}{2021}\natexlab{}.
\newblock \showarticletitle{Top-N Recommendation with Counterfactual User Preference Simulation}. In \bibinfo{booktitle}{\emph{Proceedings of the 30th ACM International Conference on Information \& Knowledge Management}} (Virtual Event, Queensland, Australia) \emph{(\bibinfo{series}{CIKM '21})}. \bibinfo{publisher}{Association for Computing Machinery}, \bibinfo{address}{New York, NY, USA}, \bibinfo{pages}{2342–2351}.
\newblock


\end{thebibliography}
